\documentstyle[amssymb,preprint,aps]{revtex}

\begin{document}
\draft
\title{Jaynes-Cummings model dynamics in Two Trapped ions \thanks{%
This work was supported by the National Natural Science Foundation of China
(19734006 and 69873015) and Chinese Academy of Science}}
\author{Hao-Sheng Zeng$^{1,2}$, Le-Man Kuang$^2$, and Ke-Lin Gao$^1$}
\address{$^1$Laboratory of Magnetic Resonance and Atomic and Molecular\\
Physics, Wuhan Institute of Physics and Mathematics, Chinese Academy of\\
Science, Wuhan 430071, People's Republic of China \\
$^2$Department of Physics, Hunan Normal University, Hunan\\
410081, People's Republic of China}
\maketitle

\begin{abstract}
We showed that in Lamb-Dicke regime and under rotating wave approximation,
the dynamical behavior of two trapped ions interacting with a laser beam
resonant to the first red side-band of center-of-mass mode can be described
by Jaynes-Cummings Model. An exact analytic solution for this kind of
Jaynes-Cummings model is presented. The results showed that quantum
collapses and revivals for the occupation of two atoms, and squeezing for
vibratic motion of center-of-mass mode existed in both two different types
of initial conditions. The maximum momentum squeezing for center-of-mass
mode in these two types of conditions are found to be 42.4\% and 43.8\%
respectively. The coherence, in the first type of initial conditions can
keeps long times, and in the second type of initial conditions, a concrete
form of coherent state is obtained, when the initial average number is very
small. 
\end{abstract}


\vskip 1cm

\narrowtext

\section{Introduction}

Jaynes-Cummings model(JCM) [1] originating from quantum optics that
describes the idealized situation of resonant interaction of an undamped
two-level system with a single, quantized harmonic oscillator turned out to
be one of the fundamental models in NMR, quantum optics, quantum electronics
and resonance physics. This model can be exactly solved and predicts some
interesting nonclassical effects both with respect to the atomic states and
the boson mode, such as atomic revivals[2] and squeezing[3]. So this model
has been studied extensively in both theoretical and experimental field.

In the field of cavity QED, experimental realization of JCM with dissipation
has been demonstrated both in microwave and in optical domain. Observation
of quantum revivals and nonclassical photon statistics with Rydberg atoms
and microwave cavities has been reported [4], and in the optical regime
vacuum Rabi splitting has been observed [5]. These experimental progress
stimulated a manifold of further theoretical studies, for example of the
quantum correlation between two interacting subsystems [6]. Moreover,
related models have been rendered successively, such as nonlinearly coupled
[7] and multiquantum JCMs [8], Raman-type [9], and three-level models [10].

Another fundamental system to exhibit, in appropriate limits, JCM dynamics
is a trapped ion interacting with laser beam. The quantized center-of-mass
motion of the ion in the trap potential plays the role of the boson mode,
which is coupled via the laser to the internal degrees of freedom. An
analogy between an undamped trapped ion and the JCM has been noted by
Blockley, Wall, and Risken [11], and by J. I. Cirac, et al., for damped
trapped ion [12]. A study for nonlinear multiquantum JCM dynamics of a
trapped ion is presented by W. Vogel and R. L. de Matos Filho [13].

In recent experiments of trapped ions, entangled states up to four ions have
been realized, and some applications, such as the proof of Bell's inequality
and decoherence-free quantum memory, have been confirmed[14]. These
improvements offered some advantageous conditions for studying properties of
multi-ion JCM. In this paper we will demonstrate that a system comprising
two trapped ions interacting with laser beam resonant to the first red
side-band of center-of-mass mode, in Lamb-Dicke regime, shows a JCM-like
dynamics. Collapses and revivals of the occupation of internal states, and
squeezing of the vibratic motion of center-of-mass mode, can be observed in
two different types of initial condition. The maximum momentum squeezing in
these two different types of initial condition are $42.4\%$ and $43.8\%$
respectively. Coherence of center-of-mass mode is also discussed in these
two types of condition. In the first type of condition, e.g. the internal
state of the two atoms being in ground and the vibratic motion of
center-of-mass mode being a coherent state, coherence can keeps long times
when the initial average number is very small. In the second type of
condition, e.g. the internal state being a superposition and motional state
being a vacuum, a concrete form of coherent state for center-of-mass mode is
obtained in another case of very small phonons. These properties for two
trapped ions are useful for quantum computation. Because quantum
manipulation involves at lest two-bit operations, and only we know the
dynamical properties of the ions very clearly, can we manipulate them well
and truly. In addition, the model for this system may be realized in present
experimental conditions[14], which will result in directly the tests of the
predictions from this paper.

This paper is organized as follows. In section II, we firstly deduce the JCM
model for $n$ trapped ions interacting with lasers in appropriate
conditions, and then present its exact solution for the case of two trapped
ions. In section III, we reveal the collapses and revivals for the
occupation of internal states of the two trapped ions in two different types
of initial condition. The coherence and squeezing for the vibratic motion of
center-of-mass mode are discussed in section IV. Finally a summary and some
conclusions are given in section V.

\section{Interaction Model}

For the sake of generality, we consider for a while $n$ ions trapped in a
linear trap which are strongly bounded in the $y$ and $z$ directions but
weakly bounded in a harmonic potential in $x$ direction. Assuming that the $%
n $ ions are illuminated simultaneously with a dispersive beam (or beams)
resonant to the first red side-band of center-of-mass mode, so that the
effects from the spectator motional modes can be neglected because of the
very large off-resonant reason[15]. The Hamiltonian for this system is

\begin{equation}
H=H_0+H_{int},
\end{equation}

\begin{equation}
H_{0}=\nu (a^{+}a+1/2)+\omega \sum_{i=1}^{n}\sigma _{iz}/2,
\end{equation}

\begin{equation}
H_{int}=\sum_{i=1}^{n}\frac{\Omega }{2}\left\{ \sigma _{+i}e^{i\eta
(a+a^{+})}e^{-i(\omega -\nu )t}+H.C.\right\}
\end{equation}

where $\nu $ and $a^{+}$ ($a$) are the frequency and ladder operators of the
center-of-mass mode. $\omega $ is the energy difference between the ground
state $\left| 0\right\rangle $ and the long-lived metastable excited state $%
\left| 1\right\rangle $ of each ion. For simplicity we further assume that
the Rabi frequency and Lamb-Dicke parameter for each ion are same. If all
ions are cooled under Lamb-Dicke limit, then we can expand the Hamiltonian
up to the first order in $\eta $. Transforming the Hamiltonian (3) to the
interaction picture with respect to $H_{0}$ and making use of rotating wave
approximation, we have,

\begin{equation}
H_{int}=-i\Omega \eta (a^{+}J_{-}-aJ_{+})
\end{equation}
where $J_{\pm }=J_{x}{\pm }iJ_{y}$ with $J_{\alpha }=\sum_{i=1}^{n}\sigma
_{i\alpha }$ ($\alpha =x,~y,~z$) being the three components of total spin
operator of $n$ ions.

E.q.(4) is namely the so-called JCM Hamiltonian for $n$ ions interacting
with a laser beam (or beams) in an ion trap. Compared with the traditional
JCM Hamiltonian, the total spin ladder operators of $n$ ions replaced the
spin ladder operators of one ion. When any one of the $n$ ions is excited,
the quantum number of $J_{z}$ component of total spin operator increases by
one, and in the meantime, the quantum number of center-of-mass mode
decreases by one. From this mode, we would naturally like to look forward to
some similar properties as shown in standard JCM.

In general case, searching for the solution of E.q.(4) is an open question.
Below we will only discuss the case that including only two trapped ions. In
this case the last term in E.q.(2) can be rewritten as $\omega J_{z}$, and
the quantum number corresponding to $J_{z}$ component only have three
values: 1, 0, -1(the value ``1'' means both the two ions are excited, ``0''
means one of the two ions is excited, and ``-1'' means both the two ions are
in ground state); Noting that in this system, the total excitations
(including electron excitations and excitations of center-of-mass mode) is a
conservation, thus we can arrange the eigenstates of Hamiltonian $H_{0}$ in
the order of total excitations,

\begin{equation}
\mid 1,~n-2\rangle ^{n},~\mid 0,~n-1\rangle ^{n},~\mid -1,~n\rangle
^{n};~\mid 1,~n-1\rangle ^{n+1},~\mid 0,~n\rangle ^{n+1},~\mid
-1,~n+1\rangle ^{n+1};\cdots
\end{equation}
Where the first numbers in each Dirac symbols represent the quantum numbers
of $J_{z}$ component and the second ones represent the excitations of
center-of-mass mode. The superscripts above each Dirac symbols represent the
total excitations.

In the Hilbert space constituted by above basis, the interaction Hamiltonian
(4) is a diagonal matrix made up of $3\times 3$ sub-matrices. Every
sub-matrix represents a subspace corresponding to a definite excitation. In
the subspace including $n$ excitations, we can get the exact propagator for
Hamiltonian (4) after a tedious deduction,

\begin{equation}
U=\left[ 
\begin{array}{lcr}
\frac{1}{2n-1} \left[(n-1) \cos{\beta} +n \right] & - \sqrt{\frac{n-1}{2n-1} 
} \sin{\beta} & \frac{\sqrt{n(n-1)} }{2n-1} \left[- \cos{\beta} +1 \right]
\\ 
\sqrt{\frac{n-1}{2n-1} } \sin{\beta} & \cos{\beta} & -\sqrt{\frac{n}{2n-1}}
\sin{\beta} \\ 
\frac{\sqrt{n(n-1)} }{2n-1}[-\cos{\beta} +1] & \sqrt{\frac{n}{2n-1}} \sin{%
\beta} & \frac{1}{2n-1}[n \cos{\beta} +(n-1)]
\end{array}
\right]
\end{equation}
where $\beta=\sqrt{2(2n-1)} \eta \Omega t$.

\section{Collapses and Revivals}

In order to study the phenomena of collapses and revivals of the two trapped
ions, imitating traditional process, we assume that initially the internal
states of the two ions are both in the ground states and the external
motional state of center-of-mass mode in a coherent state $\mid \alpha
\rangle $. Thus the whole initial state of the system is,

\begin{equation}
\rho (0)=\sum_{m^{\prime },m^{\prime \prime }=0}^{\infty }q(m^{\prime
})q^{*}(m^{\prime \prime })\mid -1~m^{\prime }\rangle \langle -1~m^{\prime
\prime }\mid
\end{equation}
with $q(m)=exp(-\frac{1}{2}\mid \alpha \mid ^{2})\frac{\alpha ^{m}}{\sqrt{m}!%
}$ are the coefficients for a coherent state expanding in number state basis.

The density operator of the internal states can be obtained by tracing over
the vibrational motion of center-of-mass mode. With above equations, we can
get the diagonal elements of the density-matrix of internal states for this
two-ion system,

\begin{equation}
\rho_{11}(t)=\sum^ \infty_{n=1} \frac{n(n-1)}{(2n-1)^2} (-\cos\beta+1)^2 p(n)
\end{equation}

\begin{equation}
\rho_{00}(t)=\sum^ \infty_{n=1} \frac{n}{2n-1} (\sin\beta)^2 p(n)
\end{equation}

\begin{equation}
\rho _{-1-1}(t)=p(0)+\sum_{n=1}^{\infty }\frac{1}{(2n-1)^{2}}(n\cos \beta
+n-1)^{2}p(n)
\end{equation}
with 
\begin{equation}
p(n)=e^{-\mid \alpha \mid ^{2}}\frac{\mid \alpha \mid ^{2n}}{n!}
\end{equation}
is the probability distribution of phonon number for the coherent state of
center-of-mass mode.

According to E.q.(8)-(11), we plotted the time evolution of diagonal
elements of density-matrix of the internal states as Fig.1, which describes
the time-evolution of the occupation-probability of electronic
energy-levels. We choose the initial average phonon number $\mid \alpha \mid
^{2}=8$, and the typical parameters $\eta =0.1$ and $\Omega =2\pi \times 500$
$kHz$ (These typical parameters are used throughout this paper, below we
will no longer narrate them repeatedly). From these curves, we can see
clearly collapses and revivals for the electronic state-occupation of the
two trapped ions. It is worthwhile to point out that, as in the traditional
standard JCM, the effect of collapses and revivals in here also increases
with the glowing of the average phonon number. However the average phonon
number may not be very large in here, because we assumed that the ions are
cooled to be under the Lamb-Dick limit. In order to observe collapses and
revivals in the case of large number of average phonons, we must look for
the solution of the system far from Lamb-Dick limit. This is a problem
awaiting to be solved.

Another possibility to obtain collapses and revivals for this system is
letting the initial state of the ions in a superposition and the vibratic
motion in a thermal state or vacuum state. This seems to be a more
reasonable assumption, because the vibratic motion of the ions seems more to
incline to a thermal state in general. For simplicity, we will suppose the
motional state being in a vacuum, because in present conditions we can
almost cool the ions to their vibratic ground state. Thus the whole initial
state of this system is,

\begin{equation}
\left| \Psi (0)\right\rangle =(a\left| 1\right\rangle +be^{i\varphi
_{1}}\left| 0\right\rangle +ce^{i\varphi _{2}}\left| -1\right\rangle
)\otimes \left| 0\right\rangle _{vib}
\end{equation}
with $a^{2}+b^{2}+c^{2}=1$.

Similar to the deduction for coherent initial state, we have the diagonal
elements of the density matrix for this kind of initial condition,

\begin{equation}
\rho _{11}=\frac{a^{2}}{9}[2+\cos (\sqrt{3}\xi t)]^{2}
\end{equation}
\begin{equation}
\rho _{00}=\frac{a^{2}}{3}\sin ^{2}(\sqrt{3}\xi t)+b^{2}\cos ^{2}(\xi t)
\end{equation}
\begin{equation}
\rho _{-1-1}=\frac{2}{9}a^{2}[1-\cos (\sqrt{3}\xi t)]^{2}+b^{2}\sin ^{2}(\xi
t)+c^{2}
\end{equation}
with $\xi =\sqrt{2}\eta \Omega $. Numerical calculation indicated that only $%
\rho _{-1-1\text{ }}$reveals clear collapse and revival. However, we may
expect more clear collapses and revivals when the number of
superposition-ion increased. Because the degree of coherence between atoms
will increase with the addition of ions. The picture for this case is
presented in fig.2.

In addition, we studied, by way of parenthesis, the behaviors of internal
state of the two ions for initial conditions that the two ions being ground
and the vibratic motion being in number, thermal and squeezing states. Some
results displayed in fig.3. The amplitude of Rabi oscillation for the
occupation of internal state $\left| \uparrow \uparrow \right\rangle $ is
almost between $0$ and $1$when the motional state be in a number state. But
when the motional state is in a thermal or a squeezing state, the
occupation-probability of $\left| \uparrow \uparrow \right\rangle $ is much
smaller.

\section{Coherence and Squeezing}

In quantum optics, coherence and squeezing are two important aspects to
describe properties of a radiative field. Coherence implies, in some sense,
the interference between different photons. And squeezing is a
phase-dependent reduction of the noise of the electric field strength (or,
of one of the field quadratures) below the vacuum level[16]. Now we borrow
these ideas to describe the properties of vibratic modes of trapped ions. In
this case, they represent the coherence between phonons and the squeezing of
position or momentum of COM mode respectively.

A quantitative measure for coherence can be defined by a fraction of the
number of coherent phonons,

\begin{equation}
\left\langle n(t)\right\rangle _{coh}=\left| \left\langle a\right\rangle
\right| ^{2}=\left| \sum_{n=0}^{\infty }\sqrt{n+1}\rho _{n(n+1)}(t)\right|
^{2}
\end{equation}
to the total phonon number,

\begin{equation}
\left\langle n(t)\right\rangle =\sum_{n=0}^{\infty }n\rho _{nn}(t).
\end{equation}

As for squeezing, we define the two quadratures of COM motion as,

\begin{eqnarray}
x &=&g(a+a^{+}) \\
p &=&ig(a-a^{+})  \nonumber
\end{eqnarray}
where $g$ may be regarded as the structure constant of a harmonic
oscillator. A quantitative condition for squeezing is that the normally
ordered variance is under zero:

\begin{equation}
\left\langle :\left( \Delta O\right) ^{2}:\right\rangle <0
\end{equation}
where $O$ means $x$ or $p$ in E.q.(18). The normally ordered variance for $x$
and $p$ can be written as,

\begin{eqnarray}
\left\langle :\left( \Delta x\right) ^{2}:\right\rangle
&=&2g^{2}\left\langle n(t)\right\rangle +2g^{2}\left\{ \sum_{n=0}^{\infty }%
\sqrt{(n+1)(n+2)}R_{e}[\rho _{n(n+2)}(t)]\right\} \\
&&-4g^{2}\left\{ \sum_{n=0}^{\infty }\sqrt{(n+1)}R_{e}[\rho
_{n(n+1)}(t)]\right\} ^{2}  \nonumber
\end{eqnarray}

\begin{eqnarray}
\left\langle :\left( \Delta p\right) ^{2}:\right\rangle
&=&2g^{2}\left\langle n(t)\right\rangle -2g^{2}\left\{ \sum_{n=0}^{\infty }%
\sqrt{(n+1)(n+2)}R_{e}[\rho _{n(n+2)}(t)]\right\} \\
&&-4g^{2}\left\{ \sum_{n=0}^{\infty }\sqrt{(n+1)}I_{e}[\rho
_{n(n+1)}(t)]\right\} .^{2}  \nonumber
\end{eqnarray}

From above two equations, we can see that in order to obtain squeezing, the
phonon density-matrix must exhibit nonvanishing off-diagonal elements $\rho
_{n(n+1)\text{ }}$ and /or $\rho _{n(n+2)}$. There may exist some different
possibilities to generate them, e.g. initially (1) internal states be in a
mono-state and vibratic motion be in a coherent state (or other states with
coherence, say, binomial state[17].); (2) motional state be in a thermal or
vacuum (means no coherence), but internal state be in coherent
superposition; (3) internal state be in a superposition and, in the same
time, vibratic motion be in a coherent state. Below we will still analysis
the two different types of initial condition discussed in section III.

For the first case, e.g. initially the internal states of the two ions are
both in ground and motional state of COM mode is in a coherent state. After
a complicate deduction, we get the density operator of vibratic motion,

\begin{eqnarray}
\rho _{lm}(t) &=&\frac{\left[ \left( l+1\right) \left( l+2\right) \left(
m+1\right) \left( m+2\right) \right] ^{1/2}}{(2l+3)(2m+3)}\left[ 1-\cos
\left( \sqrt{2l+3}\xi t\right) \right] \times  \nonumber \\
&&\left[ 1-\cos (\sqrt{2m+3}\xi t)\right] q\left( l+2\right) q^{*}\left(
m+2\right) +\left[ \frac{(l+1)(m+1)}{(2l+1)(2m+1)}\right] ^{1/2}\times 
\nonumber \\
&&\sin \left( \sqrt{2l+1}\xi t\right) \sin \left( \sqrt{2m+1}\xi t\right)
q(l+1)q^{*}(m+1)+\left[ \left( 2l-1\right) \left( 2m-1\right) \right]
^{-1}\times  \nonumber \\
&&\left[ l\cos \left( \sqrt{2l-1}\xi t\right) +l-1\right] \left[ m\cos
\left( \sqrt{2m-1}\xi t\right) +m-1\right] q(l)q^{*}(m)
\end{eqnarray}
where $q(m)$ is shown in E.q.(7).

It is easily imaginable that the average phonon number $<n(t)>$ doesn't
depend on $\varphi ,$ but increases with the augmentation of initial average
number $r^{2}$ or $<n(0)>$. Given a definite initial phonon number $r^{2}$, $%
<n(t)>$ reveals also obvious collapses and revivals as time passes (see
Fig.4). Moreover, this effects also become clear with the increasing of the
initial phonon number. In fact, the collapses and revivals of phonon number
relate to the collapses and revivals of the occupation of internal states.
During the time that the number of phonons collapses, its number almost
doesn't change. So the occupation of internal states also doesn't change,
and collapses occur at this time. In order to compare with the collapses and
revivals of the occupation of internal states in section III, we choose the
same initial condition and parameters in Fig.4 as in Fig.1.

Coherence $\gamma =\left| <a>\right| ^{2}/<n(t)>$ is also independent on $%
\varphi $, but rely on initial phonon number $<n(0)>$ and time $t$.
Coherence $\gamma $ oscillates damply on time and reduces rapidly with the
increasing of initial phonon number(see Fig.5). Thus if we want to keep the
coherence of vibratic motion for some long times, the initial average
phonons must be very small.

The squeezing of position and momentum depends on $<n(0)>$, $\varphi $ and
time $t$ simultaneously.. But from an inspection of (20)\symbol{126}(22), we
can first easily make sure that the minimums of $<:(\vartriangle x)^{2}:>$
and $<:(\vartriangle p)^{2}:>$ with respect to $\varphi $ occur at $\varphi
=0$. Then by use of numerical calculation, we find that $<:(\vartriangle
p)^{2}:>$ increases rapidly with the growing of initial average number (see
Fig.6). When $<n(0)>\geq 2.0$, the momentum-quadrature doesn't exist
squeezing. The minimum of $<:(\vartriangle p)^{2}:>$ with respect to initial
average number occurs at about $<n(0)>=0.51$. Given $\varphi $ and $<n(0)>$, 
$<:(\vartriangle p)^{2}:>$ is an oscillating function on time. In order to
see squeezing clearly, we plotted the time-evolution of $<:(\vartriangle
p)^{2}:>$ in Fig.7. As a comparison, we also plotted the time-evolution of $%
<:(\vartriangle x)^{2}:>$ in the same initial condition. The maximum
squeezing of momentum-quadrature in this type of initial condition is about $%
<:(\vartriangle p)^{2}:>=-0.424g^{2}$ which occurs at about $\varphi =0$, $%
<n(0)>=0.51$ and $t=343\mu s$. Similarly, we can determine the minimum of $%
<:(\vartriangle x)^{2}:>$ and its location.

Next, we consider the case that the initial state of the system is Eq.(12).
In this case we can deduce the density operator of motional state only
having the following nonvanishing elements,

\begin{equation}
\rho _{00}(t)=\frac{1}{9}a^{2}\left[ 2+\cos \left( \sqrt{3}\xi t\right)
\right] ^{2}+b^{2}\cos ^{2}\left( \xi t\right) +c^{2}
\end{equation}
\begin{equation}
\rho _{11}(t)=\frac{1}{3}a^{2}\sin ^{2}\left( \sqrt{3}\xi t\right)
+b^{2}\sin ^{2}\left( \xi t\right)
\end{equation}
\begin{equation}
\rho _{22}(t)=\frac{2}{9}a^{2}\left[ 1-\cos \left( \sqrt{3}\xi t\right)
\right] ^{2}
\end{equation}

\begin{equation}
\rho _{01}(t)=\rho _{10}^{*}(t)=\frac{1}{\sqrt{3}}ab\sin \left( \sqrt{3}\xi
t\right) \cos \left( \xi t\right) e^{i\varphi _{1}}+bc\sin \left( \xi
t\right) e^{i\left( \varphi _{2}-\varphi _{1}\right) }
\end{equation}

\begin{equation}
\rho _{02}(t)=\rho _{20}^{*}(t)=\frac{\sqrt{2}}{3}ac\left[ 1-\cos \left( 
\sqrt{3}\xi t\right) \right] e^{i\varphi _{2}}
\end{equation}

\begin{equation}
\rho _{12}(t)=\rho _{21}^{*}(t)=\frac{\sqrt{2}}{3}ab\left[ 1-\cos \left( 
\sqrt{3}\xi t\right) \right] \sin \left( \xi t\right) e^{i\varphi _{1}}.
\end{equation}

Similar to the above discuss, we can find average phonon number $<n(t)>$
doesn't depend on $\varphi _{1\text{ }}$and $\varphi _{2}$, and increases
with the add of $a$, $b$, and also occurs collapse and revival on time.

The coherence $\gamma $ reaches its maximum when $\cos (2\varphi
_{1}-\varphi _{2})=1$. And this maximum increases with the enhance of $c$.
In the limits that $a=0$, $b\rightarrow 0$ ( but $b\neq 0$ ) and $%
c\rightarrow 1$, coherence $\gamma \rightarrow 1$. This implies that in the
case of very small phonons( ``weak field'' ), the vibratic motion approaches
to a coherent state. In fact, in the above limits, the density-matrix
elements (23)\symbol{126}(28) can be written as, 
\begin{equation}
\rho _{00}(t)=1-\left\langle n(t)\right\rangle
\end{equation}

\begin{equation}
\rho _{11}(t)=\left\langle n(t)\right\rangle
\end{equation}

\begin{equation}
\rho _{01}(t)=\rho _{10}^{*}(t)=e^{i\left( \varphi _{2}-\varphi _{1}\right)
}\left\langle n(t)\right\rangle ^{1/2}
\end{equation}
and all other elements vanished completely ( Noting $\left\langle
n(t)\right\rangle =b^{2}\sin ^{2}\left( \xi t\right) $ ). These are just the
density-matrix elements of coherent state up to the terms of the order $%
\left| \alpha \right| $ and $\left| \alpha \right| ^{2}$(see the second
reference of [8]).

An inspection of (20)\symbol{126}(21) and (23)\symbol{126}(28), in
combination with numerical calculation suggests that for any $\varphi _{1}$,
when $\varphi _{2}=0$, $b=0$ and $c=0.96$, the squeezing of
momentum-quadrature reaches its maximum.. This maximum squeezing oscillate
with a nearly equi-amplitude on time. When $t$ satisfying $\cos (\sqrt{3}\xi
t)=-1$, the maximum squeezing $<:(\Delta p)^{2}:>=-0.438g^{2}$ for
momentum-quadrature has been obtained (see Fig.8). In fact, in the condition
that $b=0$, $a\rightarrow 0$, $c\rightarrow 1$, and $\cos (\sqrt{3}\xi t)=-1$%
, the nonvanishing density-matrix elements in (23)\symbol{126}(28) are,

\begin{equation}
\rho _{00}(t)=1-\frac{1}{2}<n(t)>
\end{equation}

\begin{equation}
\rho _{22}(t)=\frac{1}{2}<n(t)>
\end{equation}

\begin{equation}
\rho _{02}(t)=\rho _{20}^{*}(t)=\frac{1}{\sqrt{2}}e^{i\varphi _{2}}\sqrt{%
<n(t)>}
\end{equation}
where $<n(t)>=16a^{2}/9$ with $t$ determined by $\cos (\sqrt{3}\xi t)=-1$.
These are just the density-matrix elements of squeezing vacuum state (see
the second reference of [8]). Thus we also obtained a ``weak-field''
squeezing state in another group of initial parameters. In the case that $%
\cos (\sqrt{3}\xi t)\neq -1$, the momentum quadrature also has squeezing but
not reaches its maximum (see Fig 7).

It is worthwhile to point out that the coherence and squeezing discussed
above includes the rapid time dependence, which, of course, is meaningless
when slowly varying quantities are of interest. But we can slowen the
frequency of time-dependence by decreasing Rabi frequency $\Omega $.
Moreover, in current experimental conditions, dealing with problems in
microsecond is completely feasible.

\section{Summary and Conclusion}

We have studied the dynamical behaviors of two trapped ions interacting with
lasers resonant to the first red side-band of center-of-mass mode. In small
Lamb-Dick parameters and under rotating-wave approximation, the dynamics of
the system can be described by JCM-like mode. An exact analytic solution for
this type of JCM was presented. The results indicated obvious collapses and
revivals for the occupation of internal states in two different types of
initial condition. These collapses and revivals become more visible when the
average phonon number increases.

Coherence and squeezing for vibratic motion of center-of-mass mode were also
discussed in these two different types of initial condition. For one type of
initial condition that the internal states being grounds and vibratic motion
being in a coherent state, coherence can maintain long times when the
initial average phonon number is very small. A maximum squeezing $42.4\%$
for momentum-quadrature was obtained in appropriate parameters. For another
type of initial condition that the internal state being in a superposition
and the vibratic motion being vacuum, apparent expressions for coherent
state and squeezing state were acquired in suitable conditions. The maximum
squeezing for momentum-quadrature in this case is $43.8\%$.

We expect the results in this paper can bring some useful helps to other
problems, especially the quantum computation or quantum information
processing. Because the research for the dynamical properties of two trapped
ions locate completely in the field of trapped-ion quantum computation. We
also expect this paper can give rise to some other interesting discuss for
system of multiple ions, such as squeezing and coherence of vibrational
state, collapses and revivals in large Lamb-Dick parameters or for systems
containing more than two ions.

\smallskip

\smallskip

\smallskip

Fig.1. The occupations $\rho _{11}(t),~\rho _{00}(t),~\rho _{-1-1}(t)$ of
the internal state are given as functions of time. Initially the motional
state of COM mode be in a coherent state with average phonon number $\mid
\alpha \mid ^{2}=8$ and the two ions be in ground state.

Fig.2.The time evolution of $\rho _{-1-1}(t)$ for initially the internal
state be in a superposition of E.q.(12) with $a=b=c=1/\sqrt{3}$ and motional
state be in a vacuum.

Fig.3. The time evolution of $\rho _{11}(t)$ for initially the internal
states of the two ions are both in grounds and motional states are in
number, thermal and squeezing state. In three cases the average phonon
number $\left\langle n(0)\right\rangle =3$. (a) number state, (b) thermal
state, (c) squeezing state.

Fig.4. The average phonon number $<n(t)>$ is shown as a function of time for
the same initial conditions and parameters as in Fig.1.

Fig.5. The coherence $\gamma $ is shown as a function of time and initial
coherent phonon number for the same initial conditions and parameters as in
Fig.1.

Fig.6. The evolution of $<:(\vartriangle p)^{2}:>$ as a function of the
initial average number and time in given $\varphi =0$.

Fig.7. $<:(\vartriangle x)^{2}:>$ (dot line), $<:(\vartriangle p)^{2}:>$
(full line) are shown as a function of time for initial parameters $%
<n(0)>=0.51$,$\varphi =0$. When $t=343\mu s,$ momentum quadrature has a
maximum squeezing $<:(\vartriangle p)^{2}:>=-0.424g^{2}.$

Fig.8. (a). $<:(\vartriangle x)^{2}:>$ (dot line), $<:(\vartriangle p)^{2}:>$
(full line) are shown as a function of time for initial parameters $c=0.96$, 
$b=0$. When $t$ satisfies $cos(\sqrt{3}\xi t)=-1,$ momentum quadrature has a
maximum squeezing $<:(\vartriangle p)^{2}:>=-0.438g^{2}.$

\end{document}